\documentclass[pdflatex,sn-mathphys-num]{sn-jnl}

\usepackage[T1]{fontenc}
\usepackage{lmodern}
\usepackage{graphicx}%
\usepackage{multirow}%
\usepackage{amsmath,amssymb,amsfonts}%
\usepackage{amsthm}%
\usepackage[mathscr]{euscript}
\usepackage[title]{appendix}%
\usepackage{xcolor}%
\usepackage{textcomp}%
\usepackage{manyfoot}%
\usepackage{booktabs}%
\usepackage{algorithm}%
\usepackage{algorithmicx}%
\usepackage{algpseudocode}%
\usepackage{listings}%

\usepackage{amsmath}
\usepackage{siunitx}
\usepackage{textcomp} 

\begin{document}

\title[Heterogeneous Optically-Detected Spin-Acoustic Resonance in Spin Molecular Thin-film]{Heterogeneous Optically-Detected Spin-Acoustic Resonance in Solid-State Molecular Thin-film\\}


\author*[1,2]{ \sur{Kuan-Cheng Chen}}
\equalcont{These authors contributed equally to this work.}
\email{kuan-cheng.chen17@imperial.ac.uk}

\author[1,2]{\sur{Yongqiang Wen}}
\equalcont{These authors contributed equally to this work.}

\author[1,2]{\sur{Xiaotian Xu}}

\author[1,2]{\sur{Max Attwood}}

\author[1]{\sur{Jingdong Xu}}

\author[3]{\sur{Chen Fu}}

\author[1]{\sur{Sami Ramadan}}

\author[4,2]{\sur{Shang Yu}}

\author*[1]{\sur{Sandrine Heutz}}\email{s.heutz@imperial.ac.uk}

\author*[1]{\sur{Mark Oxborrow}}\email{m.oxborrow@imperial.ac.uk}

\affil[1]{\orgdiv{Department of Materials}, \orgname{Imperial College London}, \orgaddress{\street{Exhibition Road}, \city{London}, \postcode{SW7 2AZ}, \state{England}, \country{United Kingdom}}}

\affil[2]{\orgdiv{Centre for Quantum Engineering, Science and Technology (QuEST)},\\ \orgname{Imperial College London}, \orgaddress{\street{Exhibition Road}, \city{London}, \postcode{SW7 2AZ}, \state{England}, \country{United Kingdom}}}

\affil[3]{\orgdiv{College of Physics and Optoelectronic Engineering}, \orgname{Shenzhen University}, \orgaddress{\street{Street}, \city{Shenzhen}, \postcode{518060}, \country{China}}}

\affil[4]{\orgdiv{Blackett Laboratory, Department of Physics}, \orgname{Imperial College London}, \orgaddress{\street{Prince Consort Rd}, \city{London}, \postcode{SW7 2AZ}, \state{England}, \country{United Kingdom}}}


\abstract{We report an implementation of spin–acoustic resonance in pentacene thin films integrated on a high-quality-factor (high-Q) surface acoustic wave (SAW) resonator on a lithium niobate substrate. Heterogeneous optically detected spin--acoustic resonance (HODSAR) is an optically detected spin-resonance measurement in which the resonant drive is delivered mechanically by a surface acoustic wave (SAW). By leveraging the photo-excited triplet state of pentacene at room temperature, we demonstrate coherent spin manipulation via acoustic driving under zero externally applied magnetic field. The heterogeneously integrated device, referred to as HODSAR, utilizes spin--phonon coupling to achieve mechanically driven, zero-field spin resonance, opening avenues for room-temperature mechanically addressable spin control and device integration and spin-based device integration. We show that the high-Q multimode response of the SAW resonator enables spectrally selective acoustic addressing of triplet transitions near 105~MHz. Coherent control is evidenced by Rabi oscillations, with a Rabi frequency that increases linearly with the square root of the applied RF input power over the measured drive range, consistent with driven two-level dynamics under acoustic excitation. These results establish spin--acoustic resonance in a heterogeneously integrated molecular thin-film platform and provide a quantitative basis for benchmarking mechanically mediated spin control.}




\maketitle

\section{Introduction}\label{sec1}
Organic materials, when employed as spin materials, exhibit considerable promise for quantum information processing applications\cite{oxborrow2012room, shiddiq2016enhancing, gaita2019molecular,mena2024room,warner2013potential,coronado2020molecular,chen2024unlocking}. Their inherent chemically adjustable properties afford a substantially higher degree of freedom than that found in inorganic spin defects, including those prevalent in semiconductor defects\cite{coronado2020molecular}. Notably, their ability to be controlled without an externally applied magnetic field and their functionality at room temperature are attributes that greatly expand their practical utility in critical areas such as quantum sensing and quantum networks\cite{yang2000zero,yu2021molecular, bayliss2020optically,laorenza2022could}.

At present, the primary method for manipulating and analyzing these organic spin materials involves the use of Electron Paramagnetic Resonance (EPR) to investigate their spin dynamics, a technique also known as Electron Spin Resonance (ESR)\cite{wu2019unraveling}. However, ESR's dependence on electromagnetic wave processing introduces several challenges, particularly in designing high-Q microwave cavities for electromagnetic drive and detection and in achieving miniaturization of these devices. For instance, certain organic spin materials, when in a zero-field state, exhibit triplet sub-level splitting with resonance frequencies around 100 MHz in the absence of an externally applied magnetic field\cite{yang2000zero}. This necessitates the use of centimetre-sized components for 3D High-Q resonators in quantum information processing applications, posing significant integration challenges for future developments.

To address the miniaturization and integration challenges in quantum metrology, microelectromechanical systems (MEMS), particularly Surface Acoustic Wave (SAW) devices, employed as resonators in quantum sensors offer a viable solution through the effective ``quantum acoustic" manipulation of spins. Recently, SAW-based quantum acoustics have attracted attention due to their versatility in coupling with diverse quantum systems, including superconducting qubits\citep{chu2017quantum}, artificial atom\citep{gustafsson2014propagating}, nitrogen-vacancy (NV) centers in diamond\citep{maity2020coherent, macquarrie2015coherent, chen2020acoustically, golter2016coupling}, and semiconductor quantum dots\citep{schuetz2017acoustic}.  This approach enables a mechanically delivered drive channel for optically detected spin resonance -- namely spin-acoustic (phonon) coupling -- which is pertinent to quantum information processing \citep{neuman2021phononic, whiteley2019spin}. While metrology using inorganic-based sensors, such as NV-centered diamonds or SiC-based devices \citep{sohn2018controlling, whiteley2019spin, hernandez2021acoustically}, is well-established, the exploration of metrology utilizing spin dynamics of organic materials represents an extension of these approaches to organic spin systems in the field, offering new possibilities for quantum sensing and information processing  \cite{albino2019first}. This system takes advantage of heterogeneous integration to couple spins in organic materials with phonons generated by interdigital transducers (IDTs) on piezoelectric materials. This process effectively facilitates the coupling of the microwave source to SAW phonons, integrating phonon and photon-excited spin coupling, which is subsequently monitored using a single photon detector. This results in a multi-system coupling platform for quantum sensing, a system we refer to as the heterogeneous optically-detected spin-acoustic resonance (HODSAR).

In this article, we report the utilization of single-crystal-phase pentacene Pc:PTP thin-film deposited on a \(\text{LiNbO}_3\)-based High-Q Multi-Mode SAW Resonator (MMSAR) to demonstrate acoustically mediated spin control in Pc:PTP.  Pentacene has recently emerged as a promising candidate for quantum information processing and quantum metrology due to its zero-field splitting (ZFS), spin-triplet state and long-lived spin coherence properties at room temperaturet~\cite{lubert2018identifying, wu2019unraveling, wu2020invasive, oxborrow2012room,wang2024tailoring}.  Our approach effectively couples strain-induced phonons to spins in the molecular thin film without inducing magnetic fields that would affect their splitting levels.

The resonant frequencies of the \(\text{LiNbO}_3\)-based MMSAR, ranging from MHz to GHz, can be finely tuned by modifying the design of its interdigital transducers~\cite{shao2019phononic}. This microwave resonator achieves a quality factor approaching 10\,000. By integrating organic thin films as spin materials with piezoelectric materials as phonon sources, we realize spin-phonon coupling for quantum information processing. This heterogeneous integration provides a compact platform that mitigates layout constraints typically associated with microwave ESR measurements and enables mechanically mediated spin control. More broadly, it establishes a materials-compatible route for integrating molecular spin systems with wafer-level phononic and MEMS technologies relevant to quantum sensing and hybrid quantum devices. We emphasize that the contribution of this work is not to outperform conventional microwave ESR techniques in sensitivity or coherence, but to demonstrate a distinct coupling modality. Specifically, we establish spin-acoustic resonance in a molecular thin-film system under zero externally applied magnetic field, enabled by heterogeneous integration with a surface acoustic wave resonator. This mechanically addressable control pathway is compatible with MEMS fabrication and provides access to spin-manipulation regimes that are challenging to realize using traditional microwave cavities. 
\section{HODSAR}

\begin{figure}[!b]
    \centering
    \includegraphics[width=\linewidth]{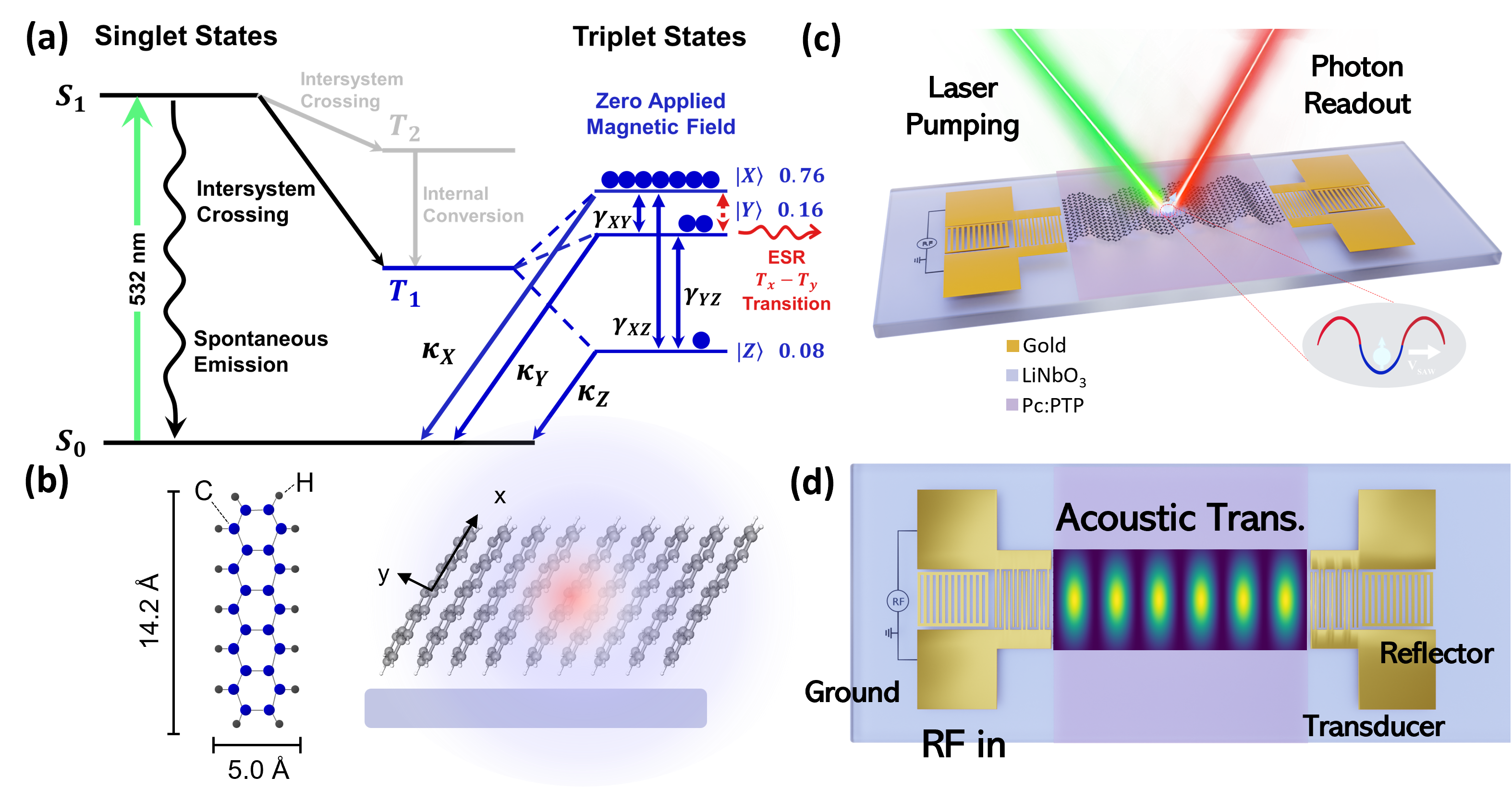}
    \caption{(a) Jablonski diagram illustrating the electronic state cycle in pentacene, responsible for spin-polarized triplet-state sublevels in pentacene:p-terphenyl. Optical excitation by a 532 nm laser promotes pentacene molecules from the ground state \( S_0 \) to the first singlet state \( S_1 \). Spin-orbit coupling mediates spin-selective intersystem crossing from the excited singlet state into the \( T_2 \) triplet manifold. The \( T_2 \) triplet states then decay non-radiatively to the \( T_1 \) triplet state, preserving the spin-polarized populations. Spin-lattice relaxation can occur between the triplet sublevels with rates \( \gamma_{XZ} \), \( \gamma_{XY} \), and \( \gamma_{YZ} \), representing the \( T_{X \rightarrow Z} \), \( T_{X \rightarrow Y} \), and \( T_{Y \rightarrow Z} \) transitions, respectively. The triplet sublevels ultimately decay back to the \( S_0 \) ground state with rates \( k_X \), \( k_Y \), \( k_Z \), completing the cycle. Resonant acoustic driving redistributes the triplet sublevel populations, which changes the spin-dependent decay pathways and is detected as a change in the measured photon counts. Our ESR readout targets the \( T_x \)-\( T_y \) transition near 105 MHz under zero externally applied magnetic field, leveraging the distinct resonance frequencies of the triplet sublevels for frequency-selective detection at zero externally applied field. (b) Structural depiction of a pentacene thin film deposited on a LiNbO\(_3\) substrate, indicating the molecular arrangement and orientation. (c) Schematic of the HODSAR device, featuring a high-Q multimode resonator that enables laser pumping and photon readout for triplet state detection. (d) Alternative schematic of the device architecture, illustrating the integration of an acoustic transducer and RF input for resonance excitation.}
    \label{fig: geo_mesh}
\end{figure}

In our experimental studies with pentacene-deposited multimode SAW resonators (MMSARs; hereafter referred to as HODSAR devices), we observe optically generated spin-triplet states in pentacene molecular thin films, following optical excitation, we rely on the same spin-selective intersystem crossing mechanism in pentacene:PTP that produces polarized triplet sublevels reported by Lubert-Perquel \textit{et al.}~\cite{lubert2018identifying}; here we differ in the drive channel, using SAW-mediated strain (rather than a microwave magnetic \(B_1\) field) to address the zero-field transition. The spin-triplet mechanism is illustrated in Fig.~\ref{fig: geo_mesh}(a): optical pumping at 532~nm excites pentacene from the singlet ground state $S_0$ to the excited singlet state $S_1$, followed by spin-selective intersystem crossing into the triplet manifold and relaxation into the triplet ground state $T_1$, generating a non-thermal (spin-polarized) population distribution among the triplet sublevels. In particular, this process produces an inverted population between the $T_x$--$T_y$ and $T_x$--$T_z$ sublevels of $T_1$, enabling optically detected spin-resonance measurements at zero externally applied magnetic field.

To acoustically address the triplet transitions, we fabricate interdigital transducers (IDTs) on a piezoelectric lithium niobate (\(\mathrm{LiNbO_3}\)) substrate, which convert an applied RF signal into a Rayleigh surface acoustic wave (SAW). In the devices studied here, the MMSAR supports a high-Q mode near 105~MHz, allowing resonant excitation of the $T_x$--$T_y$ transition of the pentacene thin film when the acoustic frequency matches the corresponding zero-field splitting. Finite-element simulations (COMSOL) following Xu \textit{et al.}~\cite{xu2022simulating} show that the oscillatory magnetic-field energy at the sample position is negligible, with a magnetic coupling efficiency of \(\eta = 5.06 \times 10^{-12}\) (see Appendix for details). This result rules out conventional microwave magnetic-dipole (\(B_1\)) driving as the origin of the observed resonance and justifies treating the experiment as mechanically driven under zero externally applied magnetic field.

\subsection{Coupling mechanism in HODSAR: strain-dominant with a possible piezoelectric electric-field pathway}
\label{sec:coupling_mechanism}

The HODSAR device couples the SAW resonator to the photo-excited triplet manifold through two channels that can, in principle, mediate the resonant response. \emph{(1) Strain-mediated coupling (dominant working hypothesis).} The Rayleigh SAW generates a time-dependent displacement field at the surface, producing a periodic strain tensor \(\varepsilon_{ij}(t)\) in the molecular thin film. This strain perturbs the effective triplet Hamiltonian by modulating the zero-field splitting (ZFS) parameters (e.g., \(D\) and \(E\)) and/or mixing the triplet sublevels, thereby enabling transitions when the acoustic frequency matches the relevant triplet splitting. In this sense, the SAW provides the resonant drive in direct analogy to ODMR, but via \(\varepsilon_{ij}(t)\) rather than an oscillatory microwave magnetic field.

\emph{(2) Piezoelectric electric-field channel (secondary possibility on \(\mathrm{LiNbO_3}\)).} Because \(\mathrm{LiNbO_3}\) is piezoelectric, the same SAW can carry an oscillating electric potential and near-field electric field above the substrate. Such fields can modulate the local molecular environment (e.g., via electrostatic Stark shifts or strain--electric-field coeffects) and may therefore indirectly affect the effective ZFS parameters. Importantly, however, this does not constitute a conventional microwave \(B_1\) magnetic-dipole driving mechanism. Consistent with this, our simulations show a vanishingly small magnetic-field contribution at the sample (\(\eta \sim 5\times10^{-12}\)), and we therefore interpret the coherent control and optically detected contrast in the main text using a strain-mediated effective Hamiltonian, while symmetry-based operator forms and electromagnetic-energy considerations are provided in the Appendix.

Motivated by this strain-mediated mechanism and by related analyses of planar spin-defect systems~\cite{udvarhelyi2023planar}, we describe the driven triplet dynamics using an effective spin--strain Hamiltonian of the form
\begin{equation}
\hat{H}_{\text{strain}}(t) = \sum_{i,j} h_{ij}\, \varepsilon_{ij}(t)\, \hat{S}_i \hat{S}_j,
\end{equation}
where $\hat{S}_i$ and $\hat{S}_j$ are the spin-1 operators along the Cartesian axes ($i, j = x, y, z$), $\varepsilon_{ij}(t)$ are the components of the time-dependent strain tensor generated by the SAW, and $h_{ij}$ are the (phenomenological) spin--strain coupling constants of the pentacene thin film. In the idealized molecular limit, symmetry considerations constrain the allowed operator combinations; in our system, a $D_{2h}$-motivated minimal model provides a physically grounded parameterization of the leading couplings, while acknowledging that local film/host environments can relax strict selection rules. The corresponding symmetry-based decomposition and operator selection rules are discussed in the Appendix.

\subsection{Optical initialization and optically detected readout}
\label{sec:optical_readout}

HODSAR uses the optically detected spin-resonance principle applied to the photo-excited triplet manifold of pentacene. A 532~nm laser pulse promotes molecules from the singlet ground state $S_0$ to the excited singlet state $S_1$, followed by spin-selective intersystem crossing that transfers population into the triplet state and produces a non-thermal (spin-polarized) distribution among the triplet sublevels. The SAW resonator provides a resonant mechanical drive that induces transitions between triplet sublevels when the acoustic frequency matches the relevant splitting, thereby redistributing their populations. Because radiative and non-radiative decay pathways from the triplet manifold depend on the sublevel population (via spin-dependent intersystem crossing and relaxation within the triplet manifold), this redistribution produces a measurable change in the detected photoluminescence (PL). In our setup, the observable is the change in photon counts collected by the SPAD in the red spectral band ($>600$~nm), reported as $\Delta \mathrm{PL}$ (or a normalized contrast) as a function of RF frequency in CW measurements or as a function of acoustic pulse duration in pulsed measurements.

A minimal rate-model picture captures the readout mechanism: optical pumping populates the triplet manifold at a rate $G$ and initializes a polarized sublevel distribution; population relaxes between sublevels with rates $\gamma_{ij}$ and decays to $S_0$ with sublevel-dependent rates $k_i$. Resonant acoustic driving adds an effective mixing (drive-induced transition) rate $\Omega_{ij}$ between the addressed sublevels, shifting the steady-state populations and hence the net PL. Operationally, this is directly analogous to ODMR, except that the resonant drive is provided by the SAW (acoustic) field rather than an oscillating microwave magnetic field.

\subsection{Coherent Acoustic Control of Molecular Thin Film}
\subsubsection{CW-HODSAR Measurement}
\label{sec:others}

\begin{figure}[!b]
    \centering
    \includegraphics[width=0.9\linewidth]{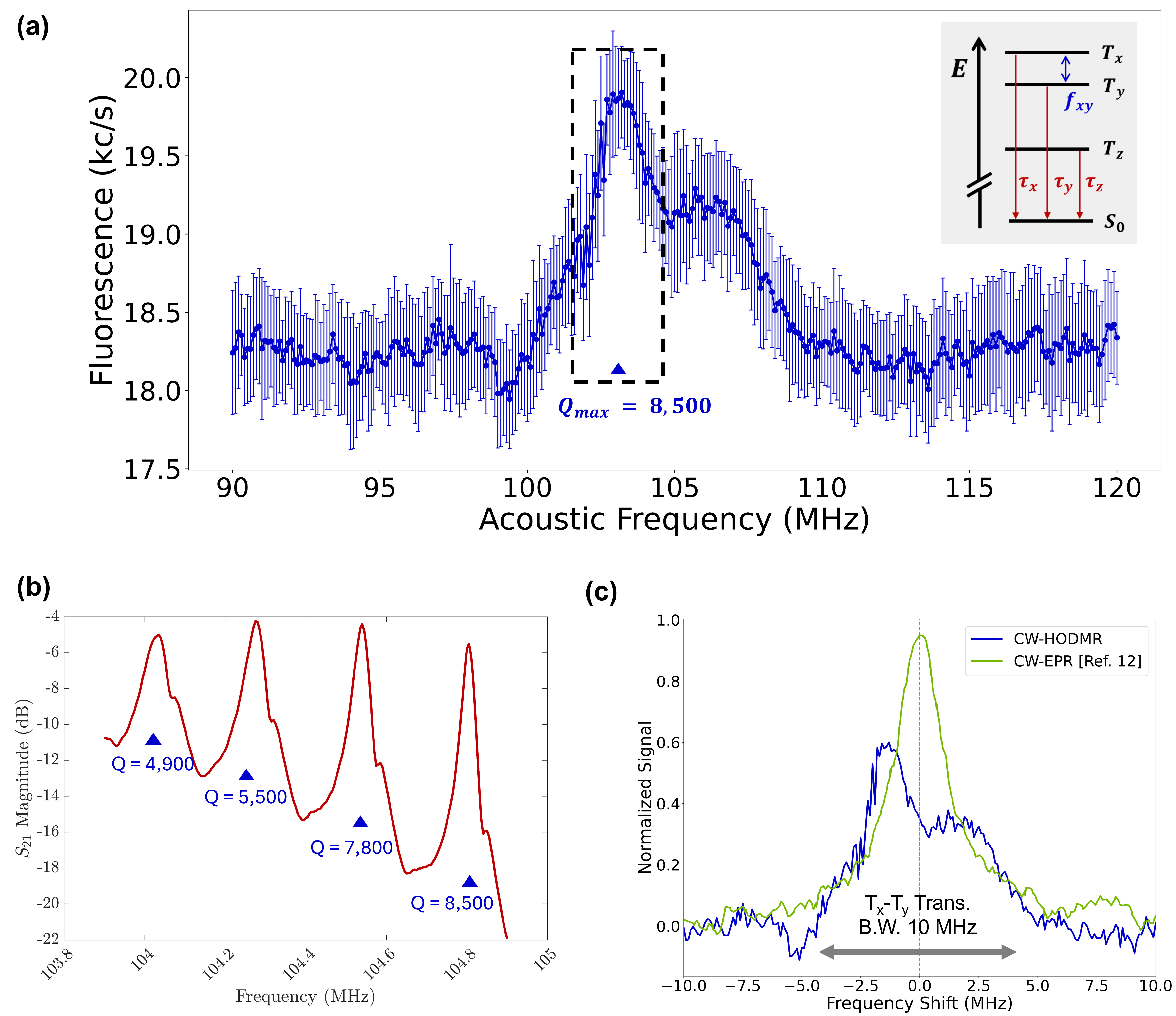}
    \caption{(a) Continuous-wave (CW) HODSAR measurements of the pentacene-deposited MMSAR device with 60 repetitions. The fluorescence data are presented with error bars indicating the standard deviation, with a resolution of 100 kHz. The dashed black box highlights the high-$Q$ resonance band around $\sim 105$ MHz. 
    Blue triangles are guide-to-the-eye markers indicating representative modes within this band.
        (b) \( S_{21} \) transmission measurements of the MMSAR device, focusing on the region within the black dashed box in (a). Four distinct high-Q resonant modes are observed in the band relevant to the \(T_x-T_y\) transition of the pentacene triplet state.
        (c) HODSAR readout of the spin-acoustic coupling, shown in comparison with continuous-wave electron paramagnetic resonance (CW-EPR) spectroscopy measurements of the hyperfine-coupled ESR. Notably, both readouts exhibit a similar bandwidth, though the line shape in (c) may be influenced by the limited high-Q multimode response of the MMSAR device across the 100-110 MHz range. The electro-mechanical properties of the MMSAR device are detailed in the Supplementary Materials.
    }
    \label{fig: cw-hodsar}
\end{figure}
To explore room-temperature acoustic control in molecular thin films, we performed continuous-wave HODSAR measurements on pentacene-deposited MMSAR devices. In CW-HODSAR, continuous optical pumping establishes a non-thermal (spin-polarized) distribution in the photo-excited triplet manifold. An applied RF signal excites a Rayleigh SAW mode, and when the SAW frequency matches a triplet splitting with zero externally applied magnetic field  (here near the \(T_x\)--\(T_y\) transition), SAW-induced strain resonantly redistributes the triplet sublevel populations. This population redistribution is converted into an optical signal through spin-dependent decay pathways and is detected as a change in fluorescence photon counts during an RF-frequency sweep. The high-Q multimode response provides spectrally selective acoustic excitation. As a result, the CW-HODSAR contrast varies with RF frequency in a manner that reflects the device transfer function across the 100--110~MHz band. Coherent control is demonstrated separately under pulsed driving via Rabi oscillations. By analyzing the fluorescence data in Fig.~\ref{fig: cw-hodsar}(a), we identify resonant features that coincide with pronounced MMSAR resonances.

We note that the measured multimode \(S_{21}\) spectrum in Fig.~\ref{fig: cw-hodsar}(b) is consistent with a \emph{coupled acoustic cavity} in which the grating reflectors provide the dominant confinement, while the periodic IDT electrode structure introduces a finite acoustic reflectivity and hence weak internal reflections within the active region. Even without an explicit IDT-only characterization in this study, the resulting fine structure can therefore be understood as a superposition of the primary mirror-defined cavity response and electrode-related Fabry--P\'erot-like contributions, which can modulate the relative amplitudes and linewidths of nearby resonances.

Comparisons with CW-EPR spectroscopy show that CW-HODSAR provides a similar bandwidth for detecting the triplet transition (Fig.~\ref{fig: cw-hodsar}(c)). However, the two measurements weight the spin ensemble differently. In CW-EPR, excitation uniformity is primarily set by the microwave \(B_1\) field distribution of the resonator/cavity and the signal is detected inductively. In HODSAR, optical pumping selects a non-equilibrium sub-ensemble within the illuminated region, while the mechanical drive is provided by SAW strain whose amplitude and phase vary spatially and depend on the acoustic mode. Consequently, the detailed line shape can differ even when probing the same underlying triplet transition. Three effects are sufficient to account for the line-shape differences: (i) multimode spectral filtering by the discrete, frequency-dependent MMSAR response in the 100--110~MHz band; (ii) spatial mode-profile weighting from non-uniform SAW strain and optical intensity across the spot; and (iii) ensemble inhomogeneous broadening arising from distributions of molecular orientation, local strain, local electric fields, and film crystallinity/grain-boundary environments. In particular, item (i) can naturally include both the dominant grating-defined cavity transfer function and smaller electrode-related reflections, consistent with the coupled-cavity picture above.

We further provide an order-of-magnitude estimate of how many molecular spins are manipulated in CW-HODSAR. Here, ``manipulated'' refers to SAW-driven transitions within the optically addressed ensemble, i.e., molecules illuminated by the laser spot and contained within the film thickness that experience resonant SAW-induced mixing of triplet sublevels. Using the microscope parameters in Methods, the diffraction-limited spot radius can be estimated as \(r \simeq 0.61\lambda/\mathrm{NA}\), giving \(r \approx 0.61\times 532~\mathrm{nm}/0.40 \approx 0.81~\mu\mathrm{m}\). To remain conservative for a 10\(\times\) objective and potential aberrations or misalignment, we take \(r \sim 0.8\text{--}3~\mu\mathrm{m}\). With film thickness \(t \sim 1~\mu\mathrm{m}\) (SEM cross-section, Fig.~\ref{fig:characterization}(a)), the addressed volume is $V \approx \pi r^2 t \sim (2.0\times 10^{-18}\text{--}2.8\times 10^{-17})~\mathrm{m}^3.$

For a 1\% Pc:PTP blend, we estimate the pentacene number density as \(n_{\mathrm{Pc}} \sim 10^{-2}(\rho/M)N_A\), using a representative organic-solid mass density \(\rho \sim 1~\mathrm{g\,cm^{-3}}\) and molar mass \(M \sim 230\text{--}280~\mathrm{g\,mol^{-1}}\), yielding \(n_{\mathrm{Pc}} \sim (2.2\text{--}2.6)\times 10^{25}~\mathrm{m^{-3}}\). The corresponding number of addressed pentacene molecules is therefore $N_{\mathrm{Pc}} \approx n_{\mathrm{Pc}}V \sim (4\times 10^{7}\text{--}7\times 10^{8}) \;\approx\; 10^{8}\ \text{(order of magnitude)}$.
Not all addressed molecules are simultaneously in the photo-excited triplet manifold: the instantaneous triplet population is a steady-state fraction \(f_T\) set by the optical pumping rate and the triplet lifetime/relaxation. Thus, the number of triplet spins contributing to a given CW-HODSAR measurement is \(N_T \approx f_T N_{\mathrm{Pc}}\), i.e., \(N_T \sim f_T\times(10^{8})\) in order of magnitude, with \(f_T \ll 1\).

\clearpage

\subsubsection{Pulsed HODSAR Measurement (Rabi Oscillation)}

\begin{figure}[!b]
    \centering
    \includegraphics[width=\linewidth]{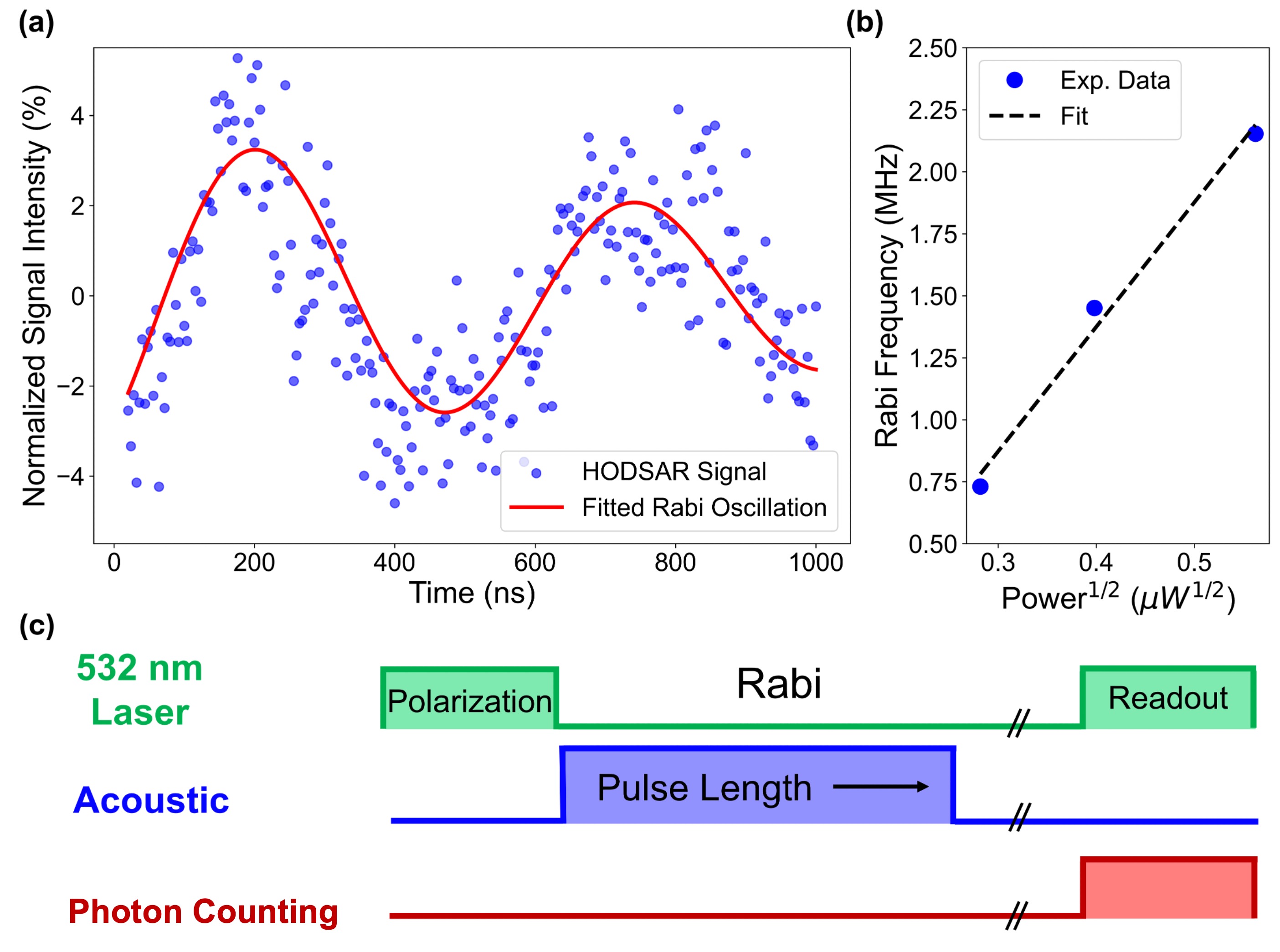}
    \caption{Characterization of Rabi oscillations in the pentacene thin-film-based HODSAR device. (a) Time-domain measurement of the HODSAR signal, showing normalized signal intensity as a function of time, with the fitted Rabi oscillation overlaid in red. (b) Rabi frequency plotted as a function of the square root of applied power, with experimental data points and a linear fit illustrating power dependence. (c) Schematic of the experimental setup for Rabi oscillation measurements, indicating the timing of laser polarization, acoustic pulse length, and photon counting readout. This characterization validates the coherence and control of Rabi oscillations in the device.}
    \label{fig: pulse-odmr}
\end{figure}

Finally, we demonstrate the pulsed coherent control of HODSAR measurements in our pentacene thin-film-based device. Drawing inspiration from the work of Dietz et al.\cite{dietz2023spin}, the pulsed HODSAR technique enables state preparation and serves as a valuable resource for entanglement generation. Utilizing a driving radio frequency of 104.5MHz—which corresponds to the highest signal intensity observed in our previous measurements—we generate acoustic resonance through IDTs. By characterizing the Rabi oscillations under pulsed excitation, we confirm the coherence and controllability of quantum states within the system. The time-domain measurements exhibit clear oscillatory behavior, indicating effective coherent manipulation of spin states. Additionally, the linear dependence of the Rabi frequency on the square root of the applied RF power confirms coherent driven dynamics under acoustic excitation. These results demonstrate controlled spin rotations in the molecular thin-film system and indicate the suitability of the platform for investigating mechanically mediated spin control protocols relevant to quantum information processing. The integration of molecular thin films with acoustic control not only offers a MEMS-compatible platform but also opens avenues for exploring complex quantum phenomena in solid-state systems.


\section{Outlook}
Our present CW-HODSAR signal originates from an optically addressed ensemble containing \(N_{\mathrm{Pc}}\sim 10^{8}\) pentacene molecules within the laser spot and film thickness, while the number of \emph{triplet} spins that participate at any instant is \(N_T \approx f_T N_{\mathrm{Pc}}\) with \(f_T\ll 1\), set by optical pumping and triplet relaxation. A conservative sensitivity benchmark is the photon-shot-noise limit, for which \(\mathrm{SNR}\simeq C\sqrt{RT}\), where \(C\) is the fractional HODSAR contrast, \(R\) the detected photon rate, and \(T\) the integration time~\cite{degen2017quantum}. For room-temperature microwave detection, the achievable sensitivity is often bounded by the effective noise temperature of the measurement chain; strategies such as cavity pre-cooling can mitigate this constraint \cite{chen2024overcoming}. In an incoherent-ensemble regime—appropriate for CW readout where the signal reflects SAW-driven redistribution of triplet sublevel populations—the contrast is expected to add approximately linearly with the number of participating spins, \(C\simeq N_T c_1\), with \(c_1\) an effective per-spin contrast determined by the local strain amplitude, spin--strain susceptibility, and collection/background conditions~\cite{degen2017quantum}. Under this assumption, extrapolating to \(N_T\sim 1\) implies a \(\sim 1/N_T\) reduction in contrast and hence a strong loss of \(\mathrm{SNR}\) at fixed \(R\) and \(T\); single-spin sensitivity would therefore require improving \(c_1\) and/or the spin-dependent photon budget rather than relying on averaging alone~\cite{degen2017quantum}.

This scaling points to three concrete levers: (i) higher optical collection efficiency and stronger background rejection to increase the usable \(R\); (ii) larger per-spin spin--acoustic transduction via stronger local strain for a given RF power (smaller acoustic mode volume, improved SAW--film overlap, and engineering that enhances ZFS strain susceptibility); and (iii) reduced addressed volume to suppress ensemble averaging and spatial inhomogeneity. Together, these directions provide a quantitative framework for assessing what would be required to move from ensemble readout toward the single-spin limit without overstating present capability~\cite{degen2017quantum}.

In this context, it is important to distinguish between an \emph{assumed} and a \emph{measured} strain susceptibility. Prior work on molecular crystals and related solid-state spin systems indicates that zero-field splittings can be modulated by strain or pressure, providing a physical basis for treating the ZFS parameters as strain dependent. While we do not directly measure an absolute ac strain susceptibility for pentacene in the present device, the observation of resonant contrast and coherent Rabi oscillations under SAW driving implies a nonzero \emph{effective} susceptibility to the driven acoustic field within the optically addressed ensemble. Quantitatively establishing this susceptibility will require independent calibration of the local strain (or stress) amplitude at the film and a controlled protocol to separate strain-driven ZFS modulation from other co-varying effects.

The same ingredients also define a pathway toward gate-level control. A natural qubit encoding is a two-level subspace of the photo-excited triplet manifold, for example \(\{|T_x\rangle,|T_y\rangle\}\) at zero applied field. Resonant SAW pulses at the corresponding splitting implement coherent rotations, with the measured Rabi frequency setting the \(X/Y\) rotation rate and controlled detuning enabling phase accumulation~\cite{maity2020coherent}; composite-pulse sequences can mitigate residual inhomogeneity in drive amplitude and local splitting~\cite{wimperis1994broadband}. Realizing single-site, single-qubit operation will additionally require spatial addressability (e.g., nano-IDTs, phononic waveguides, or mode-localized resonators), improved state-selective optical readout (e.g., microcavity-enhanced collection or near-field optics)~\cite{li2015coherent,englund2010deterministic}, and systematic coherence characterization (including \(T_2\) and driven-coherence metrics such as \(T_{1\rho}\)) to establish the operating window for acoustically mediated control in molecular thin films.

\section{Conclusions}
In this work, we demonstrate HODSAR by integrating a high-Q surface acoustic wave resonator with pentacene-deposited molecular thin films. This establishes a mechanically mediated spin-control pathway under zero externally applied magnetic field in an organic thin-film platform compatible with MEMS-scale integration and wafer-level processing. Through spin-phonon coupling, the system demonstrates spin control at room temperature with zero externally applied magnetic field using acoustic excitation. The results show that coherent manipulation of spin states can be achieved via mechanically driven transitions, providing an alternative control pathway to conventional microwave-based magnetic resonance techniques. Our findings validate the potential of organic molecular thin films as viable platforms for spin-based quantum information processing, especially in contexts where magnetic field sensitivity is a limitation. 

Moreover, the MEMS-compatibility and versatility of the SAW-based HODSAR system highlight its applicability in diverse quantum technologies, from quantum sensing to coherent spin manipulation. This study establishes a foundation for further exploration of molecular spin--phonon interactions and their applications in high-sensitivity, room-temperature quantum devices. More broadly, because heterogeneous integration can modify molecular-film morphology and texture on piezoelectric substrates, extending HODSAR to other molecular spin systems will benefit from materials screening for triplet yield, photostability, reproducible film crystallinity, and robustness to strain/electric-field perturbations. Future work will focus on enhancing the integration of organic spin systems with MEMS to expand functionality and address scalability challenges in quantum information processing.

\section{Methods}

\subsection{HODSAR Device Fabrication}
The IDTs of High-Q multi-mode SAW devices were fabricated on 128\textdegree{} YX-cut \(\text{LiNbO}_3\) wafers using a lift-off process. The wafers were sequentially cleaned with acetone, methanol, and deionized (DI) water. A photoresist (AZ5214E, Clariant Co.) was uniformly spin-coated onto the wafers and subsequently exposed to ultraviolet (UV) light through a photomask. Following exposure, the wafers underwent a reversal bake, a flood exposure, and development, employing an image reversal process to achieve a negative-tone resist profile suitable for lift-off. A 150\,nm thick gold (Au) film was deposited onto the patterned 128\textdegree{} YX \(\text{LiNbO}_3\) wafers using a thermal evaporator. The wafers were then immersed in AZ 400T photoresist stripper to remove the unwanted metal, resulting in the desired Au IDT patterns on the \(\text{LiNbO}_3\) substrates. After thorough rinsing with DI water, the wafers were diced into individual devices. For the deposition of organic molecular thin-film, pentacene (sourced from TCI UK Ltd. and purified by sublimation) and \emph{p}-terphenyl (with over 99\% purity from Alfa Aesar, further purified by zone refining) were deposited onto the SAW devices using organic molecular beam deposition (OMBD). The deposition process was carried out with a Kurt J. Lesker SPECTROS~100 system operating at a base pressure of \(1 \times 10^{-7}\)\,mbar. Quartz crystal microbalance sensors were utilized to control and monitor the deposition rates, thereby determining the composition of the thin films. The concentrations were expressed as the volume ratio of pentacene to \emph{p}-terphenyl, achieving a specific concentration of 1\% pentacene in the thin film deposited on the SAW devices. FEM simulations are made using the COMSOL 5.6 Multiphysics package to simulate the electromagnetic Effect in the HODSAR system\cite{xu2022simulating}. The detailed characterization and simulation result are shown in Supplemental Materials.

\subsection{HODSAR System Setup}
The HODSAR system setup is based on a confocal laser scanning microscope configured as follows. A 532~nm laser serves as the optical excitation source, modulated by a double-pass Isomet acousto-optic modulator (AOM). Radio Frequency (RF) driving source is provided by a Tektronix TSG4106A RF signal generator and amplified by a Minicircuits ZHL-42W+ RF power amplifier, which delivers an output power of +35.29~dBm at 10~MHz. A Minicircuits ZFSWHA-1-20+ RF switch is used to modulate the RF signal. A wheel of dichroic filters reflects the green excitation light and transmits red fluorescence above 600~nm. An Olympus UPLXAPO 10× objective lens with a numerical aperture of 0.40 delivers the excitation to the SAW resonator and collects the fluorescence. Optical detection is performed using an Excelitas single-photon avalanche diode (SPAD). The output of the SPAD is fed into a Swabian Instruments Time Tagger Ultra, which records the arrival times of photons. A Swabian Instruments Pulse Streamer generates pulse sequences to modulate the AOM, the SPAD, and the RF switch. The experimental control system is based on the Qudi software suite. The pentacene-thin-film-deposited MMSAR device is connected to the RF signal generator and positioned under the objective for signal detection. The detailed discussion is described in Supplemental Materials.

\clearpage

\section*{Acknowledgments}
In memoriam of Professor Norbert Klein, whose initial support was instrumental in the commencement of this project. The authors would like to thank Malcolm Connolly, at Imperial College London (ICL), Hao Wu, at the Beijing Institute of Technology (BIT), and Wern Ng, now at UC Berkeley, for their valuable discussions, as well as Niel Alford at ICL for his support in the project initiative. This work was supported by the U.K. Engineering and Physical Sciences Research Council through Grants Nos. EP/K037390/1 and EP/M020398/1. Additionally, K.C. acknowledges financial support from the Taiwanese Government Scholarship to Study Abroad (GSSA).

\bibliography{sn-bibliography}

\clearpage

\begin{appendices}
\section*{Appendix}


\subsection*{Effective spin--strain Hamiltonian based on \(D_{2h}\) symmetry}

Pentacene molecules integrated in thin films exhibit photo-excited triplet spin dynamics that enable optically detected spin resonance at room temperature and, in particular, operation without an externally applied magnetic field~\cite{lubert2018identifying}. To model spin--acoustic driving in our HODSAR geometry, we use an effective-spin description of the experimentally relevant triplet manifold and its coupling to SAW-induced strain, rather than a fully relativistic multi-electron derivation. In this framework, spin--orbit and spin--spin physics enter implicitly through the zero-field splitting (ZFS) parameters and through the relaxation pathways that govern optical initialization and readout.

Pentacene belongs to the $D_{2h}$ point group~\cite{stenger2009polarized}, characterized by high symmetry with multiple mirror planes, inversion centers, and two-fold rotational axes. In the ideal molecular limit this symmetry constrains the form of the effective Hamiltonian; however, in a thin-film/host environment the local symmetry can be reduced. We therefore use $D_{2h}$ considerations as a physically motivated minimal model for the leading spin--strain couplings, without assuming perfect symmetry in the device.

The total spin Hamiltonian, incorporating SS and SO interactions as well as interactions with external perturbations such as acoustic phonons, is expressed as:
\begin{equation}
\hat{H} = \hat{H}_{\text{ZFS}} + \hat{H}_{\text{SO}} + \hat{H}_{\text{strain}},
\end{equation}

For the purposes of mechanically driven spin resonance, we work within the effective triplet manifold where the dominant static term is the ZFS Hamiltonian and the drive enters through a time-dependent strain coupling. Spin--orbit coupling is not required explicitly to model the driven resonance; its primary role here is indirect, enabling spin-selective intersystem crossing and relaxation processes that underpin optical initialization and readout. Accordingly, the working Hamiltonian used below is \(\hat H(t)\approx \hat H_{\mathrm{ZFS}}+\hat H_{\mathrm{strain}}(t)\).

The zero-field splitting (ZFS) arises from magnetic dipole-dipole interactions between unpaired electron spins and is given by:
\begin{equation}
\hat{H}_{\text{ZFS}} = D \left( \hat{S}_z^2 - \frac{1}{3} S(S+1) \right) + E \left( \hat{S}_x^2 - \hat{S}_y^2 \right),
\end{equation}
where $D$ and $E$ are the axial and rhombic ZFS parameters, respectively, and $\hat{\mathbf{S}} = (\hat{S}_x, \hat{S}_y, \hat{S}_z)$ is the spin operator for $S=1$. For pentacene, experimental measurements yield $D = 1400\,\text{MHz}$ and $E = 50\,\text{MHz}$~\cite{lubert2018identifying}, reflecting significant anisotropy in the electron spin interactions due to the molecular structure. These parameters set the zero-field transition frequencies within the triplet manifold that can be driven resonantly by an acoustic mode when \(\hbar\omega_{\mathrm{SAW}}\approx |E_m-E_n|\).

The interaction with external perturbations, such as acoustic phonons, is captured by the spin-strain coupling Hamiltonian $\hat{H}_{\text{strain}}$. In a crystalline environment, strain modifies the electronic energy levels and spin interactions. In the HODSAR geometry, the SAW generates a periodic strain field \(\varepsilon_{ij}(t)\) at the film, so that \(\hat H_{\mathrm{strain}}(t)\) acts as the resonant drive (playing the role ordinarily played by an oscillatory microwave magnetic field in conventional ODMR/ESR).

In $D_{2h}$ symmetry, both the strain tensor components $\varepsilon_{ij}$ and the spin operators $\hat{S}_i$ can be classified according to irreducible representations~\cite{cotton1990chemical}. The strain components $\varepsilon_{xx}$, $\varepsilon_{yy}$, and $\varepsilon_{zz}$ transform as $A_g$ (the totally symmetric representation), while $\varepsilon_{xy}$, $\varepsilon_{xz}$, and $\varepsilon_{yz}$ transform as $B_{1g}$, $B_{2g}$, and $B_{3g}$, respectively. Similarly, the bilinear spin operator products $\hat{S}_i \hat{S}_j$ transform according to these irreducible representations.

Requiring invariance of the effective Hamiltonian under the symmetry operations (in the idealized limit) restricts the leading-order spin--strain couplings to symmetry-allowed products of \(\varepsilon_{ij}\) and bilinear spin operators~\cite{udvarhelyi2023planar}. This motivates the phenomenological form:
\begin{equation}
\begin{aligned}
\hat{H}_{\text{strain}} = & \, g_1 (\varepsilon_{xx} + \varepsilon_{yy} + \varepsilon_{zz})(\hat{S}_x^2 + \hat{S}_y^2 + \hat{S}_z^2 - S(S+1)) \\
& + g_2 (\varepsilon_{xx} - \varepsilon_{yy})(\hat{S}_x^2 - \hat{S}_y^2) + g_3 \varepsilon_{xy} (\hat{S}_x \hat{S}_y + \hat{S}_y \hat{S}_x) \\
& + g_4 \varepsilon_{xz} (\hat{S}_x \hat{S}_z + \hat{S}_z \hat{S}_x) + g_5 \varepsilon_{yz} (\hat{S}_y \hat{S}_z + \hat{S}_z \hat{S}_y),
\end{aligned}
\end{equation}
where $g_i$ are the spin-strain coupling coefficients (phenomenological constants that parameterize how SAW-induced strain modulates and mixes the effective ZFS tensor and hence the triplet energies and eigenstates).

For the $S=1$ system, the term $\hat{S}_x^2 + \hat{S}_y^2 + \hat{S}_z^2 = S(S+1) = 2$, making the first term involving $g_1$ vanish. Therefore, the spin-strain Hamiltonian simplifies to:
\begin{equation}
\begin{aligned}
\hat{H}_{\text{strain}} = & \, g_2 (\varepsilon_{xx} - \varepsilon_{yy})(\hat{S}_x^2 - \hat{S}_y^2) + g_3 \varepsilon_{xy} (\hat{S}_x \hat{S}_y + \hat{S}_y \hat{S}_x) \\
& + g_4 \varepsilon_{xz} (\hat{S}_x \hat{S}_z + \hat{S}_z \hat{S}_x) + g_5 \varepsilon_{yz} (\hat{S}_y \hat{S}_z + \hat{S}_z \hat{S}_y).
\end{aligned}
\end{equation}

With zero externally applied magnetic field and considering only the dominant terms, the total working Hamiltonian becomes:
\begin{equation}
\hat{H}(t) = D\left( \hat{S}_z^2 - \frac{1}{3} S(S+1) \right) + E\left( \hat{S}_x^2 - \hat{S}_y^2 \right) + \hat{H}_{\text{strain}}(t).
\end{equation}

Crucially, the SAW produces a time-dependent strain \(\varepsilon_{ij}(t)\) at the drive frequency \(\omega_{\mathrm{SAW}}\), so that \(\hat H_{\mathrm{strain}}(t)\) provides the oscillatory term required for resonant transitions. When \(\omega_{\mathrm{SAW}}\) matches a triplet transition, the driven dynamics redistributes the sublevel populations (CW) and enables coherent rotations (pulsed), which are subsequently converted into an optical signal through spin-dependent decay pathways.

Acoustic phonons induce strain in the lattice, modulating the electronic environment and, consequently, the spin interactions. In a symmetry-based picture, the dominant coupling channels are those where the strain symmetry matches that of the relevant bilinear spin operators, yielding selection rules in the idealized limit; in textured films these rules can be relaxed, but the same operator structure remains a useful guide for identifying the leading drive terms.

Phonons transforming as $B_{1g}$ (shear strain in the $xy$-plane) couple to $(\hat{S}_x^2 - \hat{S}_y^2)$ and $(\hat{S}_x \hat{S}_y + \hat{S}_y \hat{S}_x)$. Phonons transforming as $B_{2g}$ and $B_{3g}$ couple to $(\hat{S}_x \hat{S}_z + \hat{S}_z \hat{S}_x)$ and $(\hat{S}_y \hat{S}_z + \hat{S}_z \hat{S}_y)$, respectively.

Spin-acoustic resonance occurs when the energy splitting between spin sublevels matches the energy of the acoustic phonons, allowing for coherent transitions driven by lattice vibrations without external magnetic fields. The resonance condition is given by:
\begin{equation}
\hbar \omega_{\text{phonon}} = E_{\text{spin}},
\end{equation}
where $E_{\text{spin}}$ is the energy difference between spin states within the ZFS-split triplet manifold. In our experiment the observable is an optically detected contrast (change in photon counts / PL) arising from this driven redistribution of triplet populations.

Quantitative strain susceptibilities and sensing figures of merit require independent calibration of the local strain field and are left for future work.

\subsection*{Simulation for Electromagnetic Effect in the System}

\begin{figure}[!b]
    \centering
    \includegraphics[width=0.85\linewidth]{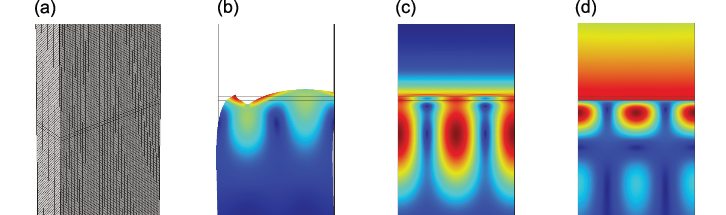}
    \caption{(a) Schematic illustration of the simulation geometry with the used mesh showing the pentacene sample on the lithium niobate substrate. The corresponding simulation results are depicted in the subsequent three figures: (a) the displacement field, (b) the electric field, and (c) the magnetic field around the pentacene sample. Deformation is added to the displacement field for visualization purposes.}
    \label{fig:sim_res}
\end{figure}

The effective-Hamiltonian picture used in the main text assumes that the resonant drive is delivered mechanically (via SAW-induced strain) rather than by an unintended oscillatory magnetic field. We therefore complement the spin--strain model with finite-element simulations that quantify the relative electromagnetic and electromechanical energy densities at the sample location, allowing us to bound any magnetic-drive contribution.

We employed COMSOL Multiphysics\cite{oxborrow2007traceable}, a finite-element analysis software, to simulate the electromagnetic fields around the pentacene sample. Our primary focus was on the magnetic field generated by the driven SAW mode to rule out the possibility of a conventional microwave magnetic-dipole (spin--photon) drive mechanism. We adopted the method introduced by Xu \textit{et al.}\cite{xu2022simulating}.

The simulation targeted the Rayleigh mode and concentrated on the area covered by the sample. As illustrated in Fig.~\ref{fig:sim_res}(a), the computational domain was divided into two regions by the surface of the lithium niobate substrate: an internal space representing the substrate and an external space representing the region above it. Two \textit{Block}s (COMSOL terminology is italicized throughout), each with dimensions of 32 µm (width), 10 µm (depth), and 95 µm (height), were used to model these spaces. The substantial height was necessary to accommodate the transverse evanescent fields.

The external space was further partitioned to represent the pentacene sample and the surrounding air. Specifically, a \textit{Block} with dimensions of 32 µm $\times$ 10 µm $\times$ 1 µm represented the pentacene sample. The mesh strategy is depicted in Fig.~\ref{fig:sim_res}(a). A \textit{Free Quad} mesh was applied to the surfaces to ensure accurate field resolution.

The simulation results are presented in Fig.~\ref{fig:sim_res}, showing (b) the displacement field, (c) the electric field, and (d) the magnetic field around the pentacene sample. Deformation is added to the displacement field for visualization purposes.

To quantify the coupling coefficient $\eta$, we integrated the energy densities to calculate the total energy of each field component. The coupling coefficient is defined as

\begin{align}
    \eta = \frac{E^{\text{ext}}_{\text{mag}}}{E^{\text{int}}_{\text{ela}} + E^{\text{int}}_{\text{kin}} + E^{\text{int}}_{\text{ele}} + E^{\text{int}}_{\text{mag}} +  E^{\text{ext}}_{\text{ele}} + E^{\text{ext}}_{\text{mag}}} = 5.07 \times 10^{-12},
\end{align}

where $E^{\text{ext}}_{\text{mag}}$ is the magnetic energy in the external region, $E^{\text{int}}_{\text{ela}}$ is the elastic energy, $E^{\text{int}}_{\text{kin}}$ is the kinetic energy, $E^{\text{int}}_{\text{ele}}$ is the electric energy, and $E^{\text{int}}_{\text{mag}}$ is the magnetic energy in the internal region.

The extracted $\eta$ demonstrates that the oscillatory magnetic-field contribution at the sample is vanishingly small compared with the elastic and electric energy stored in the SAW mode. Consequently, the observed resonant response is most consistently attributed to strain-mediated modulation/mixing of the triplet Hamiltonian (spin--strain coupling), rather than a conventional microwave magnetic-dipole drive.

\clearpage

\subsection*{Proccess Flow of HODSAR Device Fabrication}

\begin{figure}[htpb]
    \centering
    \includegraphics[width=0.85\linewidth]{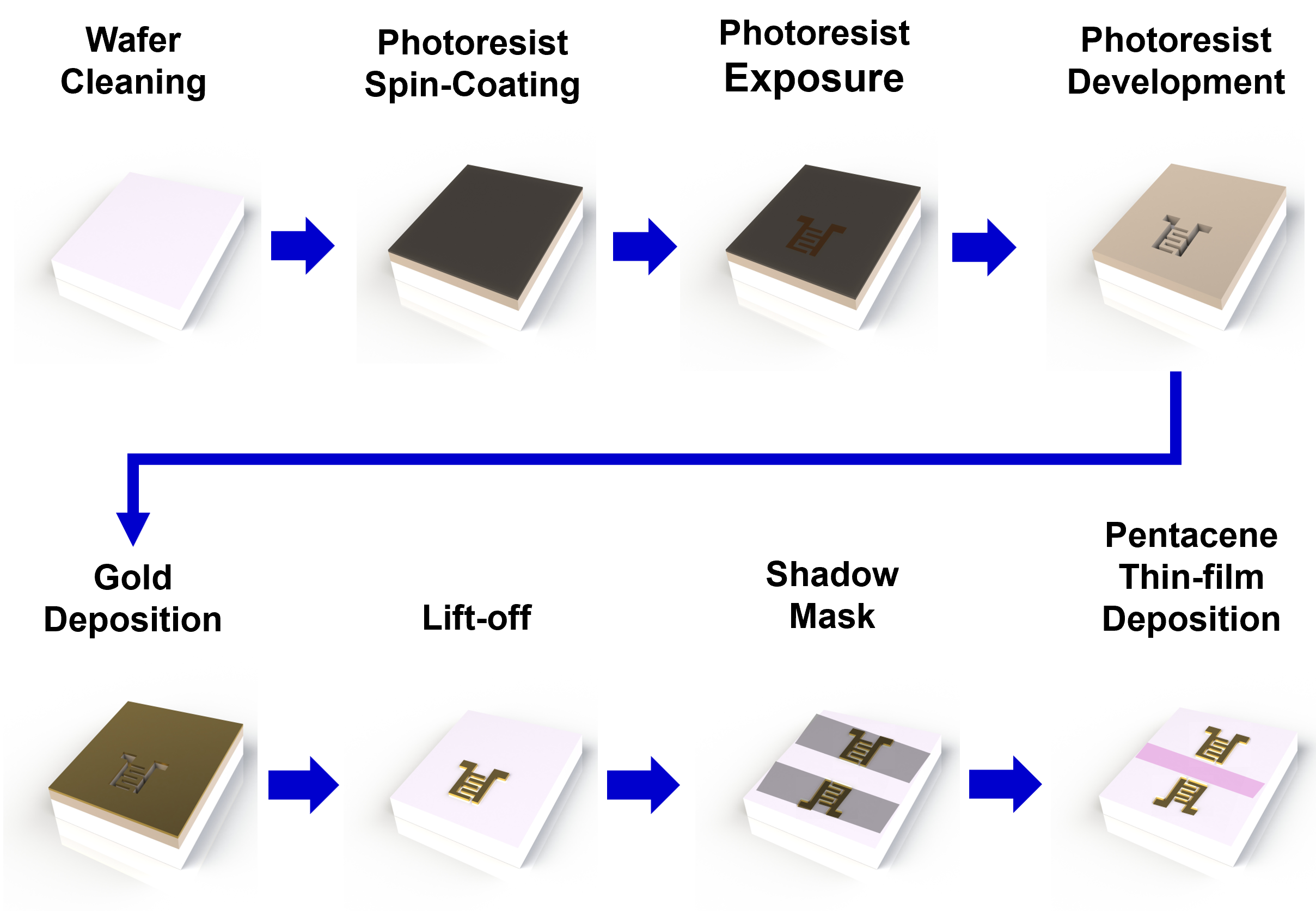}
    \caption{Fabrication process for pentacene thin-film-based high-Q organic dielectric surface acoustic resonator devices (as mentioned as HODSAR devices) with gold electrodes. The procedure starts with wafer cleaning, followed by photoresist spin-coating. Photoresist exposure and development define the electrode pattern, followed by gold deposition and lift-off to establish the electrode structure. A shadow mask is then applied to enable selective deposition of the pentacene thin film on targeted areas. }
    \label{fig: fabrication}
\end{figure}

\clearpage

\subsection*{HODSAR Setup }
The HODSAR setup is based on a confocal laser scanning microscope that consists of a 532 nm laser as the optical excitation source, a double-pass Isomet acoustic-optic modulator (AOM), an RF signal generator (Tektronix TSG4106A) as the RF drive source and a RF power amplifier (Minicircuits ZHL-42w+) that has an amplification of +35.29 dBm at 10 MHz, a RF switch (Minicircuits ZFSWHA-1-20+)that modulate the RF, a wheel of dichroic filters that reflect the green excitation light and transmit the red fluorescence above 600 nm, an Olympus UPLXAPO 10X Objective with a numerical aperture of 0.40 that delivers the excitation to the MMSAR devices and collect the fluorescence, an Excelitas single-photon avalanche diode (SPAD) as the optical detection. 
The output of the SPAD is fed into a Swabian Instruments Time Tagger Ultra that tags the arrival of photons.
A Swabian Instruments Pulse Streamer is used to generate pulse sequences to modulate the AOM, the SPAD and the RF switch.
Aside the hardware part, the experiment control system is based on Qudi system suite \cite{binder2017qudi}.

\begin{figure}[htpb]
    \centering
    \includegraphics[width=1.0\linewidth]{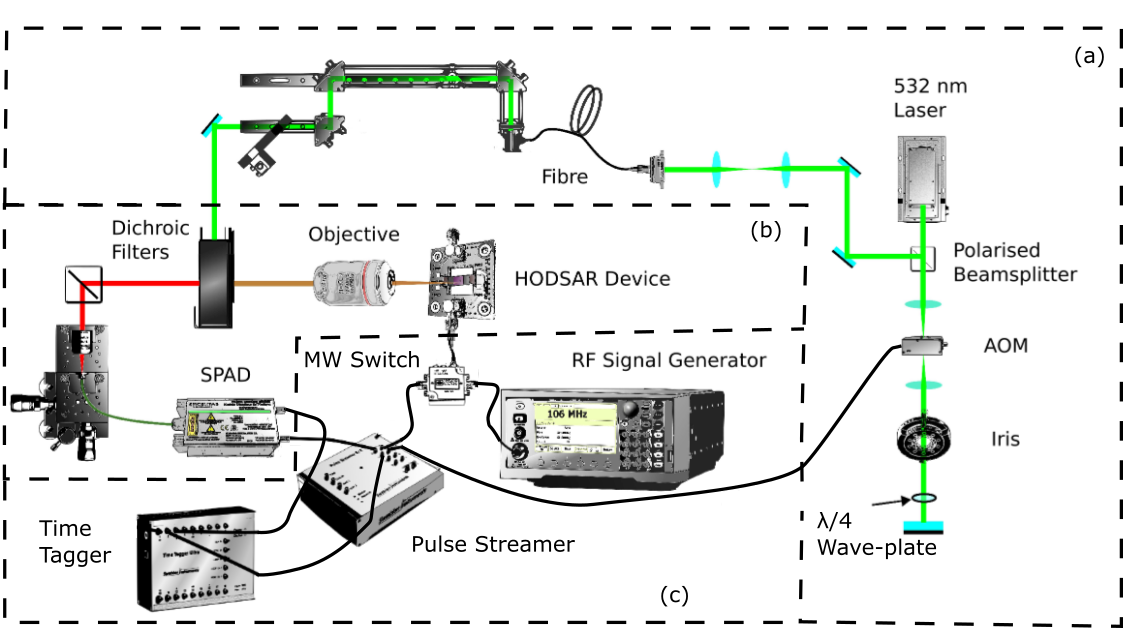}
    \caption{A schematic diagram of the CW and pulsed HODMR measurement setup. (a) The HODSAR device is exited by a 532 nm laser that is modulated by a double-passed AOM. (b) The excitation laser is reflected towards the HODSAR device and the fluorescence is filted by a 600 nm long-pass dihroic mirror and transmitted to the SPAD. (c) The Pulse Streamer is used to generate the pulse sequences for the RF switch, the SPAD and the AOM, the Time Tagger counts the singals from the SPAD and the synchronisation signal from the Pulse Streamer. The RF signal generator generates a RF signal and it is modulated by the switched and delivered to the MMSAR device.}
    \label{fig: hodsar-setup}
\end{figure}

\clearpage



\subsection*{High-Q MMSAR Electro-Mechanical Characteristics}

\begin{figure}[!b]
    \centering
    \includegraphics[width=0.85\linewidth]{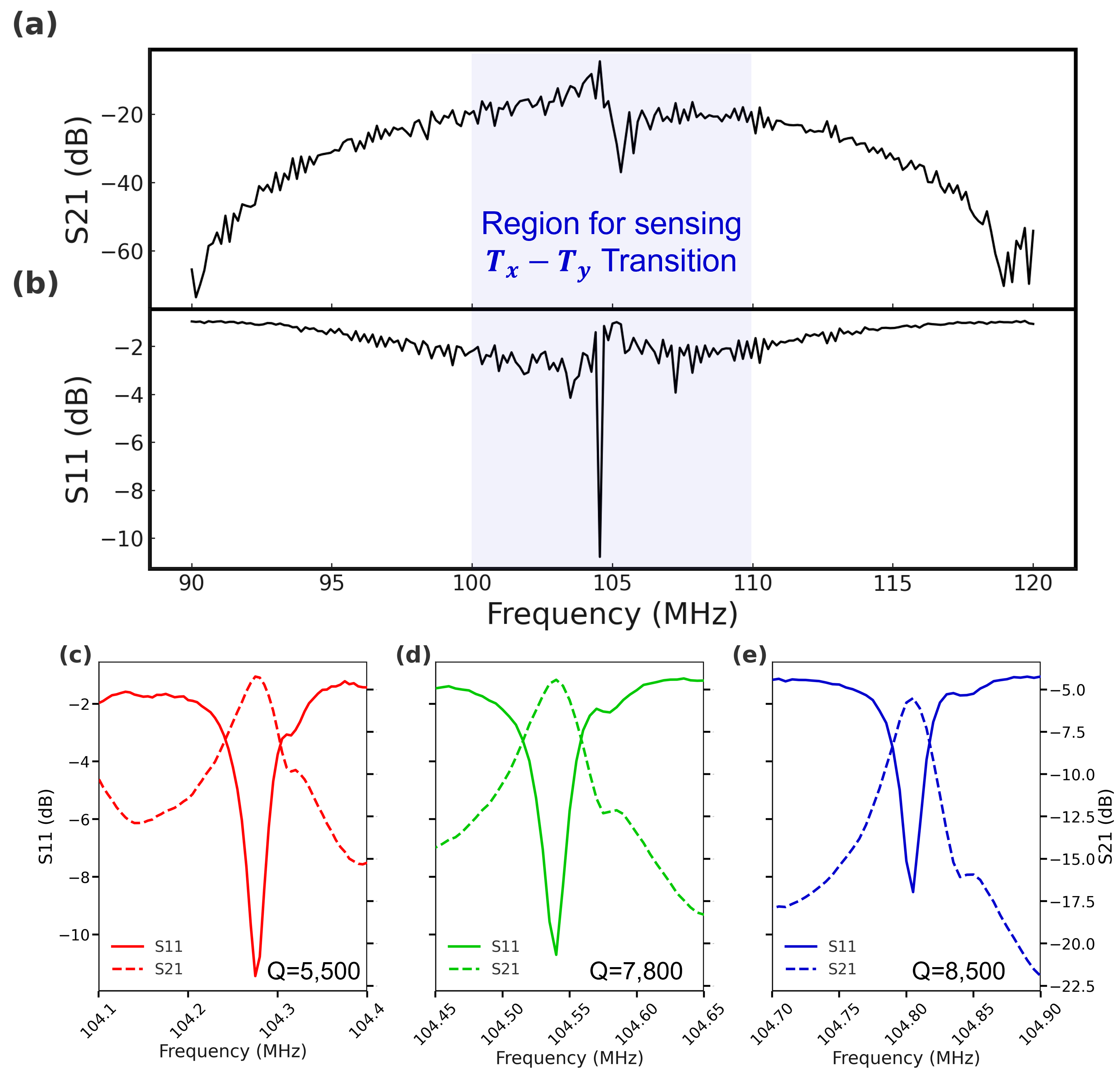}
    \caption{Two-port electrical measurements of the HODSAR device are presented. Panels (a) and (b) show the transmission (S$_{21}$) and reflection (S$_{11}$) spectra, respectively, over a wide frequency range, highlighting the region of interest for sensing the \(T_\text{x}-T_\text{y}\) transition of photoexcited pentacene thin films. The blue-shaded region indicates the spectral range where these transitions occur. Panels (c), (d), and (e) display the highest Q-factor responses observed within the blue-shaded region, with corresponding Q-factors (calculated using the Q-circle method\cite{kajfez1984q}) of 5,500, 7,800, and 8,500. These measurements confirm multiple narrow resonances with extracted quality factors in the frequency band relevant to the \(T_x-T_y\) transition.}
    \label{fig:s2p}
\end{figure}

\begin{figure}[htpb]
    \centering
    \includegraphics[width=0.6\linewidth]{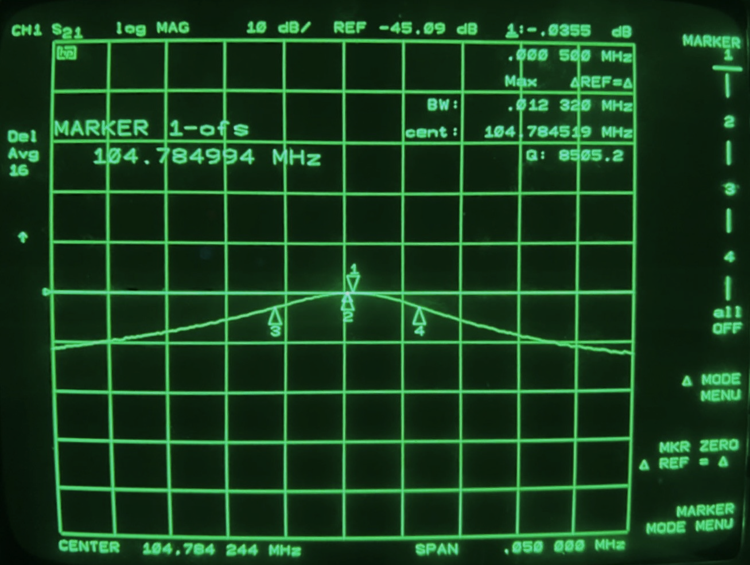}
    \caption{The HP8753C vector network analyzer (VNA) measurement illustrates the method used to determine the Q-factor values in Fig. S4(c), with a representative Q-factor of 8,505.2 at a frequency centered at 104.8MHz.}
    \label{fig:vna}
\end{figure}

The electromechanical characteristics of the HODSAR device were assessed using two-port vector-network-analyzer measurements, recording both the transmission (\(S_{21}\)) and reflection (\(S_{11}\)) spectra over a broad frequency range (Fig.~\ref{fig:s2p} and Fig.~\ref{fig:vna}). The primary goal of this characterization is to identify high-\(Q\) acoustic resonances that provide spectrally selective SAW excitation in the vicinity of the pentacene zero-field splitting relevant to the \(T_x\!-\!T_y\) transition. Within this band, we observe multiple narrow resonances; their quality factors (extracted using the Q-circle method) reach \(Q \approx 5.5\times 10^{3}\), \(7.8\times 10^{3}\), and \(8.5\times 10^{3}\), confirming that the MMSAR supports low-loss, frequency-selective modes suitable for acoustically driven spin control.

The measured multimode structure in \(S_{21}\) is consistent with a coupled acoustic cavity formed primarily by the grating reflectors, with additional internal reflections introduced by the periodic IDT electrode arrays. In addition to providing electromechanical transduction, the IDTs impose an acoustic impedance modulation and therefore a finite reflectivity, so that weak Fabry--P\'erot-like contributions can coexist with the dominant mirror-defined cavity response\cite{plessky2000coupling}. The observed fine structure and mode-to-mode variations in linewidth and peak height near 105~MHz can thus be understood as a superposition of the primary cavity resonances and smaller electrode-related reflections. We note that an IDT-only response was not isolated in this study; future device iterations will separately characterize IDT-only and grating-cavity contributions to quantitatively disentangle these effects.

\subsection*{Pentacene Thin-Film Characterization}

\begin{figure}[!b]
    \centering
    \includegraphics[width=\linewidth]{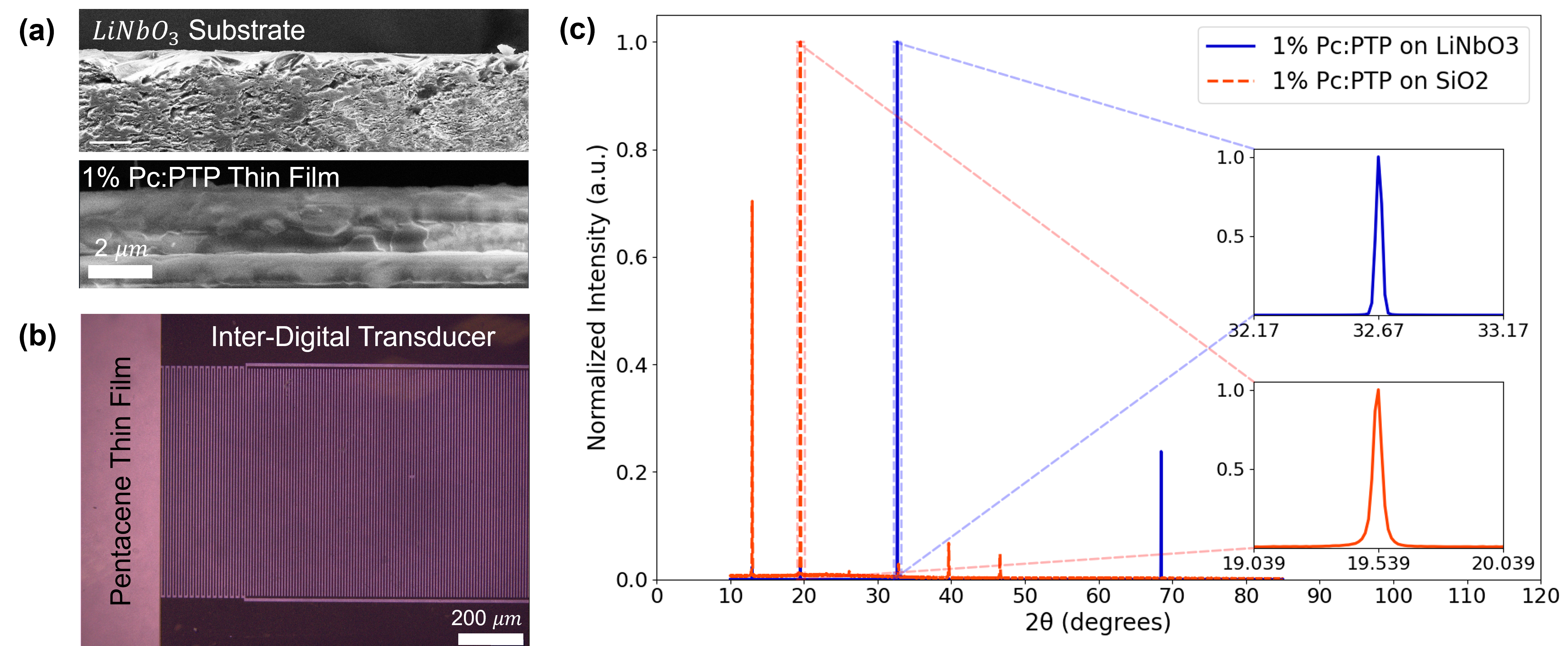}
    \caption{(a) SEM cross-sectional image of the MMSAR device, showing the 1\% Pc:PTP thin film layer on a LiNbO\textsubscript{3} substrate with a 2 $\mu$m scale bar. (b) Optical microscope top view of the HODSAR device, highlighting the pentacene thin-film region and IDTs with a 200 $\mu$m scale bar. (c) XRD pattern comparison of the 1\% Pc:PTP thin film on LiNbO\textsubscript{3} (solid blue line) and SiO\textsubscript{2} (dashed red line) substrates, demonstrating structural differences with expanded views of peak regions around 32.67° and 19.54°.}
    \label{fig:characterization}
\end{figure}

In this paper, we benchmark the growth of pentacene thin films on a 128$^\circ$ Y-cut LiNbO$_3$ substrate, comparing our results to those reported by Lubert \textit{et al.}~\cite{lubert2018identifying} on SiO$_2$ substrates. The organic molecular beam deposition tool, deposition recipe, and conditions were maintained consistently across both substrate types. Characterization results shown in Fig.~\ref{fig:characterization}(a) indicate that the 1\% Pc:PTP blend films exhibit a grain size of approximately 0.7~$\mu$m on LiNbO$_3$, which is smaller than the 1~$\mu$m grain size observed on SiO$_2$ substrates~\cite{lubert2018identifying}. This substrate-dependent change in grain size is consistent with differences in nucleation density and coalescence kinetics during early-stage growth.

X-ray diffraction (XRD) analysis shown in Fig.~\ref{fig:characterization}(c) shows clear substrate-dependent differences in diffraction peak positions and relative intensities, indicating that the Pc:PTP film adopts different preferred orientations and/or microstructural textures on LiNbO$_3$ compared with SiO$_2$. Specifically, we observe a dominant reflection near 30$^\circ$ 2$\theta$ on LiNbO$_3$ that is not apparent in the SiO$_2$ reference, while the SiO$_2$ film exhibits prominent peaks near 6$^\circ$ and 18$^\circ$ 2$\theta$ consistent with the texture reported previously~\cite{lubert2018identifying}. Taken together, these observations support the conclusion that the substrate influences the film texture and crystalline packing, without requiring an epitaxial relationship.

Several non-exclusive mechanisms can account for these differences. LiNbO$_3$ and SiO$_2$ present distinct surface energies, surface terminations, and electrostatic boundary conditions, which can modify molecular adsorption, diffusion lengths, and the critical nucleus size during OMBD growth. In addition, LiNbO$_3$ can impose anisotropic near-surface strain fields (and, under certain conditions, piezoelectric surface potentials) that may bias the orientation distribution and stabilize different textures during crystallization. Accordingly, we interpret the additional reflections and the altered intensity pattern on LiNbO$_3$ as signatures of substrate-dependent texture and/or a mixed-orientation microstructure, rather than evidence for a new phase or improved electronic performance. A definitive assignment of phase fractions and orientation relationships would require complementary measurements (e.g., pole figures/GIXRD, rocking curves, and/or cross-sectional TEM), which are beyond the scope of the present work.

These substrate-dependent morphology and texture changes are likely to be generic for other molecular thin films integrated onto piezoelectric SAW platforms, and they can translate directly into a broader distribution of local environments (e.g., orientation- and strain-dependent ZFS parameters) that contributes to inhomogeneous linewidth and device-to-device variability. Accordingly, extending HODSAR to other molecular systems will benefit from screening candidate molecules and host matrices for (i) high triplet yield under optical pumping, (ii) photostability under repeated illumination, (iii) reproducible crystallinity when grown on piezoelectric substrates, and (iv) robustness of the triplet Hamiltonian to strain and electric-field perturbations. Establishing these materials criteria alongside the device transfer function will help separate intrinsic spin-physics limits from substrate- and growth-induced microstructure effects.

\end{appendices}

\end{document}